# Hysteresis and nucleation in condensed matter

Yuri Mnyukh
*76 Peggy Lane, Farmington, CT, USA,  e-mail: yuri@mnyukh.com*
(Dated: March 11, 2011)

The physical origin of hysteresis in condensed matter had not been previously identified. The current "science of hysteresis" is useful, but limited by phenomenological modeling. This article fills the void by revealing the exclusive cause of the hysteresis in structural, ferromagnetic and ferroelectric phase transitions, as well as upon magnetization in magnetic fields and polarization in electric fields. This exclusive cause is *nucleation lags*. The lags are inevitable due to the nucleation specifics, far from the classical "random fluctuation" model.  A major assumption  that spin orientation is determined by the orientation of its carrier explains why ferromagnetic transitions and magnetization in magnetic fields materialize by structural rearrangements at interfaces, as well as why magnetization by "rotation" is impossible. Formation of the structural and ferromagnetic hysteresis loops is considered in detail.

## 1. Conventional science: phenomenological modeling

Hysteresis, lags of a process whichever direction it goes, is widely present in the condensed matter, both as an adverse effect and a phenomenon useful in technological applications. In the conventional literature its investigation has been limited by mathematical descriptions of its manifestations. The three-volume set *The Science of Hysteresis* [1] is an example of such approach. A more appropriate title would be "The Science of Hysteresis Modeling" due to absence in its 2160 pages of anything about the true physical nature of the phenomenon. From practical point of view the theoretical modeling of hysteresis is useful. But wouldn't it be better to do it already understanding its physical nature?  The purpose of this article is to fill the void.

## 2. "What causes magnetic hysteresis?"

Probably, the most consequential is *magnetic hysteresis*. Its cause has not been found by the current theory. The sixteen authors of [2] think that it will be beneficial to understand it, asking "What causes magnetic hysteresis?". They argue, correctly, that magnetic hysteresis is fundamental to magnetic storage technologies and a cornerstone to the present information age. They found that all the "beautiful theories of magnetic hysteresis based on random microscopic disorder" failed to explain their data. Their answer to their own question was: "New advances in our fundamental understanding of magnetic hysteresis are needed".

## 3. The key to hysteresis is nucleation

In the meantime, the physical origin of hysteresis in crystal structure rearrangements was disclosed years ago. It revealed itself as an inherent part of the molecular mechanism of crystal phase transitions during their systematical investigations [3-16], summarized and extended to the magnetic hysteresis in [17]. Hysteresis is absent in second-order phase transitions by definition [18], but not a single undisputed example of such transition is found [17, 19]. The fact that first-order phase transitions are realized by *nucleation and growth* provides the direction where to look for the cause of hysteresis. The n*ucleation* is the key to its explanation. Nucleation in solid-state rearrangements will be outlined here first. It is far from the conventional "random fluctuation" model.

## 4. The cause of hysteresis as seen in optical microscope

In the present context the following experiments could be regarded "*experimental* hysteresis modeling", as opposed to the usual theoretical modeling. Solid-state phase transitions in a number of organic crystals were investigated by direct observation in optical microscope equipped with a heating/cooling stage.[4,5]. Small (~1 mm) transparent single crystals of *p*-dichlorobenzene (PDB) with the temperature $T_0 = 30.8$ °C, such that the free energy $F$ of its H (above $T_0$) and L (below $T_0$) phases are equal, $F_H(T_0) = F_L(T_0)$, were the most convenient object. Every crystal was subjected to a slow heating or/and cooling. Here are some results. Fig. 1 is a photograph of phase transition in one of the heating experiments. Phase transitions start from *nucleation* after $T_0$ has been passed. Nucleation is always *heterogeneous*, located at the crystal defects. The actual temperature of phase transitions does not - and cannot - coincide with $T_0$, considering that no reason exists for the transition to go in any direction when the free energies are equal. In other words, the temperature $T_0$, usually called "phase transition temperature" (and sometimes even "critical temperature") is the temperature where phase transitions cannot occur. But the possibility to occur at any other temperatures, except a small region around $T_0$, is theoretically unlimited.. Heterogeneous nucleation requires a finite energy for activation, which makes *threshold* nucleation lags inevitable. If $T_n$ is actual temperature of nucleation, the minimum (threshold) overheating in PDB was $\Delta T_n = T_n - T_0 = \sim 1.9\ ^oC$. Landau and Lifshitz [18] were incorrect by stating that overheating or overcooling is *possible* in first-order phase transitions. Hysteresis is inherently *inevitable*, even though it can be very small (see section 8 below). In general, better crystals exhibit wider *hysteresis*. An extreme case is the observation of melting at 53.2 $^oC$ of a PDB crystal, still in its L-phase, being "too perfect" to contain even a single suitable defect to serve as the H-nucleation site. This also illustrates absence of a homogeneous nucleation.

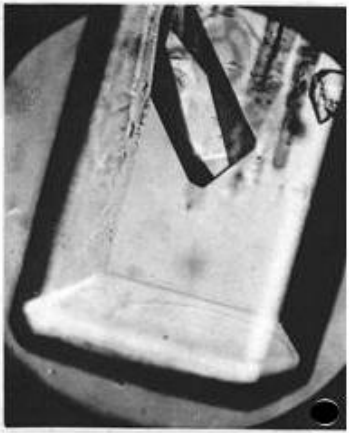

Fig. 1. An example of phase transition in a transparent single crystal of *p*-dichlorobenzene. It is a crystal growth of well-faceted single crystal of the new (higher-temperature) phase within the initial (lower-temperature) phase. The transition started from a visible crystal defect. There is no rational crystallographic orientation relationship between the initial and new phases..

### 5. "Non-classical" peculiarities of nucleation

At this point, the above-described reality has brought about essential information on the origin of hysteresis. *Nucleation* is its exclusive cause. More is hidden in the peculiarities of nucleation [14]. Only *optimum microcavities* - conglomerations of vacancies, and not any other kind of defects, serve as the nucleation sites. The nucleation is not a random successful fluctuation: it is *predetermined*. Special experiments have revealed that every potential nucleation site contains a "pre-coded" individual temperature of its activation $T_n$. In the repeat experiments the $T_n$ was the same as long as the phase transition was initiated by the same nucleation site. The $T_n$ was different in different crystals. Generally, if the temperature is slowly rising, and there are several potential nucleation sites, the one of lowest temperature $T_n$ would be actually activated.

### 6. Range of transition

The "jumps" of physical properties in phase transitions are never instant. Upon heating or cooling they always spread over a temperature range, narrow or wide, exhibiting a "sigmoid" curve (Fig. 2). It is usual to (erroneously) take the inflection point of the curve as the "transition temperature" or even as the "critical point". Phase transitions over "wide" temperature range are called *diffuse*. The "diffuseness" is not the manifestation of a specific transition mechanism. It results from the non-simultaneous nucleation in different particles, or parts, of the specimen. Any transition in a powder or polycrystalline specimen is "diffuse", the variation being only how much. The width of a transition range is not a fixed value, being a characteristic of the particular crystal imperfection, rather than an inherent property of the substance. For instance, the range of transition will be sharply different for a single crystal and the powder made from it. Concluding, (a) range of transition is range of *nucleation*, (b) it is affected by the sample condition / preparation, (c) it lies entirely outside $T_0$: above it upon heating and below it upon cooling.

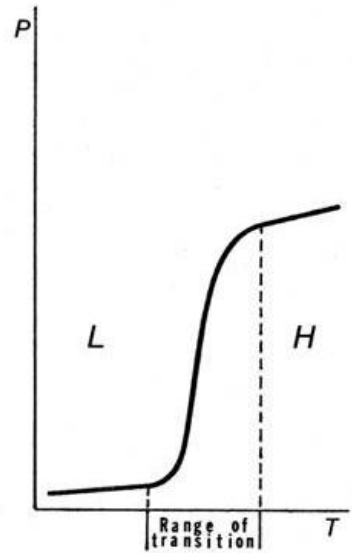

Fig. 2. Typical "sigmoid" plot of a physical property P upon heating through a phase transition.

### 7. Hysteresis loop of a phase transition

The AE and GD curves in Fig. 3 represent mass fraction $m_H$ of the H-phase. The sigmoid curve P (T) in Fig. 2 simply delineates the mass ratio of the phases in the two-phase range. In a proper scale it becomes the right part of the hysteresis loop (AE in Fig. 3). Existence of the right part necessarily means that the left part (GD) would be found in the reverse run. It is important to note that the phase transition can be reversed only after T is lowered below $T_0$, and even further to exceed a certain threshold range of stability, to activate an L-nucleus. The hysteresis loop in a small single crystal particle is rectangular. The sigmoid shape of the plots AE and GD is for the systems of many particles. It is indicative of two factors acting in opposite directions as the temperature changes. They are: (1) increase in the number of suitable nucleation sites per unit mass, and (2) decrease of the mass of the original phase. The former factor dominates in the initial stage, and the latter in the final stage of the phase transition.

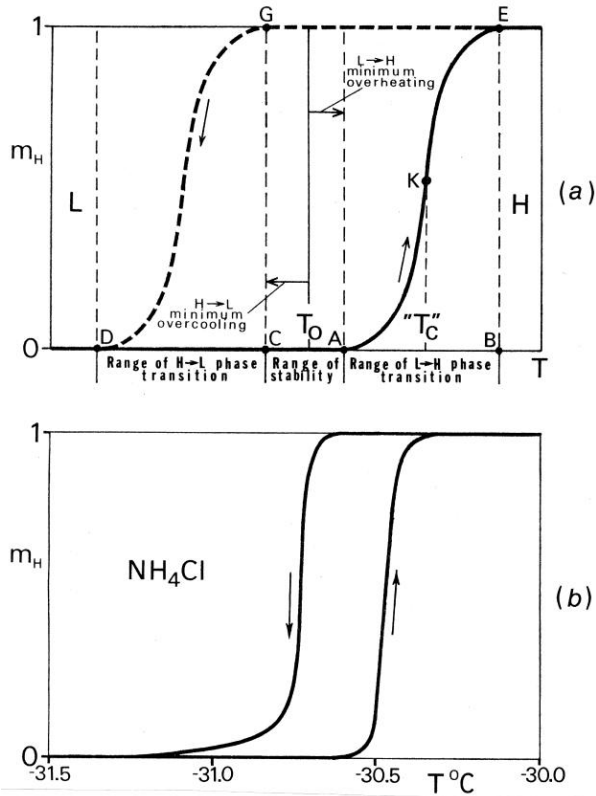

Fig. 3. Hysteresis loop of a solid-state phase transition ($m_H$ is mass fraction of H phase).
(a) (Schematic) "Sigmoid" curves AE and GD, each representing $m_H$ in the heterophase (L+H) temperature range of transition. Together they form a hysteresis loop DAEGD. Range of stability CA consists of two threshold lags too small to activate nucleation within. The inflection point K is not a "critical point" (or "Curie point"). It marks the temperature of the maximum number of activated nucleation sites.
(b) (Experimental) The hysteresis loop in $NH_4Cl$ (Dinichert [20]).

## 8. When hysteresis is small

There are circumstances when hysteresis in solid-state phase transitions (and any other structural rearrangements in solids for that matter) can be especially small. Two such cases will be outlined here, but they are presented in greater detail in [13] and [17 (Sec. 2.8 and 2.9.2 - 2.9.5)].

One is nucleation in layered crystals. A layered structure has strongly bounded, energetically advantageous two-dimensional (2-D) units - molecular layers, while the interlayer interaction is weak. Change from one polymorph to the other mainly involves the mode of layer stacking. Real layered crystals always have numerous defects resulted from imprecise layer stacking. Most of these defects are minute microcavities in the form of wedge-like interlayer cracks concentrated at the crystal faces. In such a microcavity there is always a point where the gap has the optimum width for nucleation. There the molecular relocation from one side of interface to the other occurs with no steric hindrance and, at the same time, with the aid of attraction from the opposite wall. In view of a close structural similarity of the layers in the two polymorphs, this nucleation will be *epitaxial*. There is a simple answer to why the temperature hysteresis $\Delta T_n$ in *epitaxial* phase transitions is small. Due to the abundance of wedge-like microcracks, there is no shortage in the nucleation sites; at that, the presence of a substrate of almost identical surface structure acts like a "seed". As a result, only small overheating or overcooling is required to initiate phase transition. Without a scrupulous verification, these phase transitions may seem "instantaneous", "without a hysteresis", "cooperative", "displacive", "second-order", etc.

Besides the hysteresis caused by formation of a 3-D nucleus to initiate the structural rearrangement, there is another type of hysteresis, much smaller, related to propagation of the immerged interfaces. It has been shown that advancement of an interface in normal direction requires a 2-D nucleation. The 2-D nuclei form heterogeneously as well. There is a significant difference in their function. While only a single 3-D nucleus from an *optimum microcavity* is needed to start a transition, a sufficient concentration of appropriate defects is required to keep the interface moving. These defects are also microcavities, but smaller, although not just individual vacancies. They are *vacancy aggregations*. One such defect can act as a 2-D nucleus only once. It disappears when the current layer is completed and another one is needed to build next layer. Existence of two kinds of nucleation - and the associated two-level hysteresis - means that a phase transition, ones started by a 3-D nucleus, may continue at a lower overheating/overcooling provided the temperature is kept outside the 2-D nucleation threshold. Another consequence relates to the exact value of $T_0$. It is not enough to find it as the temperature when the interface does not move in any direction. The two phases are not in dynamic equilibrium at their interface. The $T_0$ can be determined only approximately as being within the temperature range consisting of the positive and negative 2-D nucleation threshold lags.

## 9. Magnetization hysteresis from the new fundamentals of ferromagnetism

The term "magnetic hysteresis" actually implies hysteresis of *magnetization*. The attempts to explain magnetic hysteresis will be unproductive until the physics of magnetization is understood. The current interpretation of magnetization process is not valid. It rests on the Heisenberg's obsolete theory of ferromagnetism assuming the existence of extremely strong *electron exchange interaction*. But even the initial verifications of that theory had to prevent its acceptance: *the theory produced a wrong sign of the exchange forces*. Despite this grave defect, it has been taken for granted. But Feynman [21] remained skeptical, noting that "even if we *did* get the right sign, we would still have the question: why is a piece of lodestone in the ground magnetized?", and concluding that "this physics of ours is a lot of fakery." Later on, the sign problem was examined again [22] and found unavoidable. It was suggested that the "neglect of the sign may hide important physics." The new physics of ferromagnetism was put forward in 2001in the book [17]. Here are its main principles:
1. The Heisenberg's strong electron correlation in ferromagnetics does not exist. Contribution of the magnetic dipole interaction to the total crystal free energy is small as compared to that of crystal bonding. A ferromagnetic crystal is stable due to its low *total* free energy in spite of a possible small destabilizing effect of the magnetic interaction.
2. The phase transitions and other structure changes are realized by nucleation in specific crystal defects and rearrangement at interfaces in all instances, including all ferromagnetic and ferroelectric phase transitions.
3. Orientation of a spin is a unique characteristic of its atomic carrier. Therefore, orientation of spins in a crystal lattice is set by the orientation of the particles (atoms, molecules) in the crystal; any re-magnetization (during or without phase change) proceeds exclusively by the nucleation-and-growth rearrangement of the crystal.

The new fundamentals of ferromagnetism instantly explain both a magnetization and its hysteresis. The common notions "*switching*" and "*reversal*" imply instant change of spin orientation in the crystal lattice. They are inconsistent with the possibility of hysteresis and could not explain why experimentally estimated ultimate speed of "magnetization switching" in single-domain particles turned out orders of magnitude lower than theoretically predicted. The solution is simple. Spin orientation is a fixed property of every particular crystal structure. *Any spin reorientation results from a reconstruction of the crystal itself*. It can be activated by a change of temperature, pressure, or application of external magnetic field. In any case it is a relatively slow process involving nucleation and propagation of interfaces. And it is a subject of the nucleation lags as presented above. *Magnetic hysteresis is nothing but a structural hysteresis* both in ferromagnetic phase transitions and in magnetization of domain systems. The answer to "What causes magnetic hysteresis?" is: nucleation lags of the underlying structural rearrangement.

## 10. Ferromagnetic hysteresis loops

Ferromagnetic hysteresis loops of magnetization *M* in external alternating magnetic fields *H* are a prominent feature of ferromagnetic materials. It is accepted as a fact that motion of domain boundaries is the main mechanism of (re)magnetization, and that the hysteresis is lags in that motion and lags in formation of new domains. But the question why magnetization is localized on the domain boundaries has not been raised. The cause of the lags has not been identified. Besides, it is erroneously accepted that magnetization can also occur by "rotation" without motion of the domain boundaries. A possible relationship between the magnetic hysteresis loops $M = f(H)$ and the "structural" ones $P = f(T)$ in solid-state phase transitions (Fig. 2 and 3) has not attracted due attention. These failures were rooted in the interpretation of the lags as those of a *magnetic rearrangement in the crystal structure*, rather than a *rearrangement of the crystal structure* itself.

The similarity between the magnetic and structural hysteresis loops is not accidental: they represent the same process of phase transition by nucleation and growth under the action of a variable. In the polymorphic transitions (section 7) the variable was T, while in ferrmagnetics it is *H*. It should be remembered, though, that the former is a scalar, while the latter is a vector, but this difference is not of basic importance. Changing T transfers a matter across the line in the phase diagram that separates its areas of stability and instability. It is the unstable state of the matter that is the driving force of the phase transition -. nucleation and growth of new phase. Similarly, the application of a magnetic field *H* to a ferromagnet makes it unstable and initiates a nucleation-and-growth rearrangement at the domain boundaries toward a state of a lower energy where spins better aligned with the *H* direction. In other words, magnetization by application of magnetic field is a solid-state phase transition, and the only way phase transitions realized is by nucleation and restructuring at the interfaces. Even though the two crystal "phases" are the same, the resultant one appears in a different crystallographic and spin orientation.

## 11. Ferromagnetic rectangular hysteresis loop

When analyzing ferromagnetic hysteresis loops one has to take into account whether
- the sample is a single crystal, polydomain crystal or polycrystal,
- the magnetic field *H* is applied in the "easy" or any other direction,
- the *H* strength is sufficient to magnetize the sample to saturation,
- the loop is quasi-stationary, or it is recorded in a high frequency alternating field.

The actual shape of the hysteresis loops varies depending on these conditions, but, like their phase transition counterparts, they can be entirely accounted for in terms of the structural categories of nucleation and growth.

In the classical experiments by Sixtus and Tonks (S & T), described in a number of sources ([*e. g.*, [23,24]), the experimental arrangement allowed investigating magnetic hysteresis loop free of complicating side effects. There the sample - ferromagnetic wire - was turned to a single crystal magnetized to saturation $M_S$, while the magnetostriction effects and internal strains were eliminated. As a result, the sample exhibited a rectangular hysteresis loop as in Fig. 4. Reducing *H'* to zero and even applying negative field $H'' < H_n$, where $H_n$ is the field necessary to start remagnetization, does not affect the $M = M_S$ value. The line A-B is horizontal. An additional applied negative field $\Delta H$, so that $|H''+\Delta H| > H_n$ would trigger formation of a nucleus with 180°-reversed magnetization, followed by the fast propagation of a domain interface over the whole sample. The line BC is vertical. Speed of the interface propagation was measured by different authors; it varied depending on the sample and on the strength of the magnetic field. The maximum speed was well below of what can be expected from a "magnetization wave". The magnetic field $H_n$ needed to create the nucleus was called "starting field" ("nucleation field" would be a better name). A somewhat weaker field, called "critical field", was sufficient to keep the interface moving.

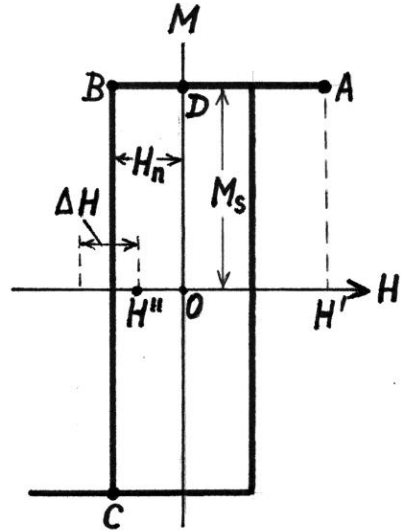

Fig. 4. Ferromagnetic rectangular hysteresis loop. See text for symbols and description.

Setting aside the shortcomings of the interpretation of the S & T experiments in [23], we will look at the subject in terms of the structural nucleation-growth concept. Remagnetization is not just a "wave of magnetization reversal": change in the $M_S$ direction occurs by the rearrangement of crystal structure at the interface. One may argue that the crystal structures on sides of the domain interface are the same and, by definition, are not different phases. This argument is valid when the variable affecting the crystal free energy is a scalar, such as temperature or pressure, but magnetic field H is a vector. The free energies of two structurally identical domains differently oriented in the magnetic field are not the same, which is the driving force of structural rearrangement at the domain interfaces.

The next point to clarify is the reason why $M_S$, achieved at the strongest positive magnetic field *H'*, remains unchanged after *H'* is reduced to zero and even farther into the negative side (horizontal line A-D-B in Fig.4). The crystal structure is stable over the region A-D, indeed. In the region D-B, on the contrary, the sample is in the unstable state, since the direction of its magnetization is opposite to *H*. It remains quasi-stable simply because no structural change can occur without nucleation, i.e., until the negative field is sufficiently strong to increase the instability to the point when a structural nucleus of the opposite magnetization appears. This "starting field" $H_n$ has the same function as the overheating / overcooling in initiating temperature phase transitions. Considering that the $H_n$ value is "pre-coded" in a structural defect, it is not exactly reproducible in different samples. This behavior is no different from the temperature solid-state phase transitions: an energetically unstable phase remains quasi-stable until conditions for the formation of a 3-D nucleus are provided.

Since the domain interface motion was regarded in literature a "wave of magnetization reversal", there was a problem to explain why speed of this wave is so low. What really takes place at the domain interfaces is not a "wave", but a structural rearrangement Structural phase transitions provide answer to the questions of why some excessive magnetic field ("critical field"), lower than a "starting field", is still required to keep the domain interface moving. The molecular mechanism of structural rearrangement at the domain interfaces is the same as in the structural phase transitions described above. It involves 3-D nucleation to start and 2-D nucleation to continue the process of magnetization. The nucleation lags are the cause of the ferromagnetic and ferroelectric hysteresis loops.

## 12. Typical ferromagnetic hysteresis loop

The rectangular hysteresis loop (Fig. 4) is at the basis of all ferromagnetic hysteresis loops. Only nucleation and growth are involved in its formation. The conditions for the rectangular loop to form are: a monodomain crystal, elimination of the magnetostriction adverse effect, a sufficiently strong magnetic field applied parallel to the direction of spontaneous magnetization, quasi-stationary recording. The shape of a typical quasi-stationary ferromagnetic hysteresis loops, like in Fig. 5 (only its upper part is shown), always deviates from being rectangular to one or another degree. The overall cause for the "typical" loop to not be rectangular is, evidently, that at least some of those conditions are not satisfied. As a rule, sufficient relevant information does not accompany real hysteresis loops, if at all. These loops are usually related to polycrystals, the fact being given little or no attention, much less properly taken into account.

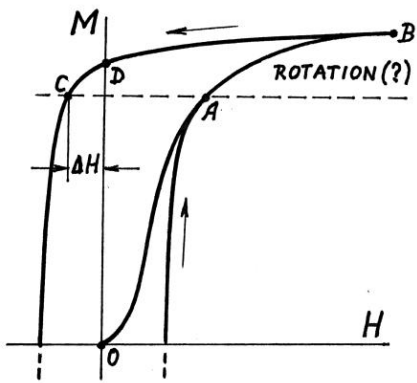

Fig. 5. A typical remagnetization hysteresis loop (its lower part is omitted). See text for explanation of its particulars. Its part over the dotted line was erroneously claimed to be due to spin rotation in the structure.

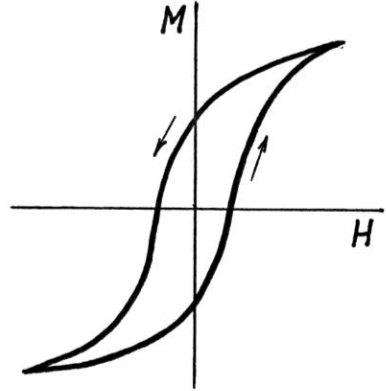

Fig. 6 The type of a ferromagnetic hysteresis loop frequently used to illustrate the phenomenon. Such loops are not quasi-stationary and therefore do not fit for analyzing their shape.

Not infrequently the illustrative hysteresis loops look like the one in Fig. 6. They have such a shape when being not quasi-stationary due to recording in fast alternating fields, instead of slow recording or point-by-point. If the applied field $H$ changes too fast, the domain interfaces do not have enough time to reach their quasi-stable positions corresponding to the $H$ amplitude. The shape of such a loop depends on the frequency of the alternating field. Besides, the relaxation time of the internal strains caused by the magnetostriction is too short and is a function of the frequency as well.

Only quasi-stationary loops are fit to be analyzed. The detailed analysis [17, (Sec. 4.13.3)] goes beyond the scope of this article. The conventional interpretation by Bozorth [23] was inadequate. A major issue is whether magnetization can occur by a "domain rotation". This magnetization process is claimed to take place in the part of loop marked "rotation (?)" in Fig. 5. According to Bozorth, at point A the magnetization stage owing to motion of the domain boundaries is completed; the magnetic moments of all domains in the sample became uniformly aligned (magnetically saturated) in the "easy" direction of the crystal; farther magnetization in the $H$ direction (from A to B) proceeds by a "reversible rotation" of the magnetic moments from the "easy" direction into the direction of the applied magnetic field $H$. It cannot be so, however. The polycrystalline material was treated as if it was a polydomain "single" crystal. Even more indicative is that the alleged rotation of $M_S$ from the "easy" direction by the magnetic field can only be elastic, because the crystal forces will try to return $M_S$ to the "easy" direction. In other words, magnetization by "rotation" has to be reversible, and Bozorth called it as such. But the actual process is not reversible, as evident from the fact that the curve B→ C does not follow B→ A. There are a number of reasonable causes, discussed in [17], to account for the ascending M from A to B. Impossibility of spin rotation in the crystal lattice is a major point in the new fundamentals of ferromagnetism [17]. It is rooted in the fact that spin directions are fixed in their particles and therefore fixed in a given crystal structure [25].

The magnetization B→ D shows that a major portion of the structural rearrangements that occurred on the way from A to B is retained, but there is some regression causing the observed slope. Whichever processes led to the magnetization A→ B, it was accompanied by accumulation of internal strains opposing this magnetization. A subsequent decrease in the $H$ strength allows strains to relax by means of structural readjustments at the expense of the magnetization. By annealing the sample under the conditions marked by point B the strains can be eliminated. In this way the curve B→ D→ C can be flattened, even made horizontal. Some $M$ decline over B →D → C is not a remagnetization yet. It begins only after $H$ changes its sign to the opposite and exceeds a certain threshold -$\Delta H$ to initiate nucleation of the oppositely oriented domains. But the sample is still polycrystalline and, contrary to a rectangular loop where a single nucleation act caused a propagation of the domain interface over the whole sample, this time one nucleation act affects only one crystal grain. The "starting fields" $H_n$ are different in different grains. The process would not proceed without |$H$| increases. Still, this is the most effective magnetization phase (|$dM/dH$| = max), ending at the point equivalent to point A.

A common misconception should be dispelled regarding the role of crystal defects in a magnetization process and formation of hysteresis loops. The defects were always considered only as an obstacle to the motion of domain boundaries. In fact, their role is twofold. In a defect-free crystal neither motion, nor even formation of the boundary is possible. We can imagine a ferromagnet exhibiting a very high coercive force because its crystal structure is "too perfect". Indeed, the shortage of adequate defects for nucleation in very small ferromagnetic particles requires very strong fields for their remagnetization. On the other hand, different kinds of crystal defects that are not suitable to serve as heterogeneous nucleation sites may hamper propagation of domain boundaries.

## 13. Ferroelectric hysteresis and hysteresis loops

Almost the entire description of ferromagnetic hysteresis and hysteresis loops is directly applicable to the ferroelectric hysteresis and its loops of repolarization in electric fields $E$. Only two points should be noted..

The *orientational* polarization (*i. e*., by "rotation") is not possible as well. The difference is that here it is self-evident. The orientation of the electric dipoles in polar dielectrics is an element of their crystal structure. Reorientation of the dipoles in a ferroelectric crystal can occur only by rearrangements of crystal structure at the domain interfaces. Ferroelectrics can be polarized to saturation $P_S$ only in a direction determined by the crystal structure, and not in the arbitrarily chosen direction of the applied field $E$. Achieving $P_S$ in polycrystalline ferroelectrics in any $E$ direction is conceivable only through growth of the grains that happened to be polarized in the $E$ direction.

Another feature is the *induced polarization*. While spontaneous magnetization $M_S$ by itself does not appreciably depend on $H$, a spontaneous polarization $P_S$ depends on $E$ to some extent. It is due to the fact that the two electric charges of a ferroelectric dipole are spatially separated in the crystal unit cell. The induced polarization adds to the polarization caused by the structural domain rearrangements and is noticeably present in the hysteresis loops. Specifically, the saturation polarization $P_S$ ($E$) continues to grow even after all the dipoles are parallel. The induced polarization is strictly reversible and has therefore nothing to do with hysteresis, even though it somewhat affects the hysteresis loop shape. For example, a ferroelectric hysteresis loop cannot be quite rectangular.

## 14. The conclusion

Where nucleation is - there is hysteresis.

## References

[1] *The Science of Hysteresis: 3-volume set,* Ed. G. Bertotti and I Meyergoyz, Academic Press, 2006.
[2] M.S. Pierce, C.R. Buechler *et al.*, *Phis. Rev*. **B 75**, 144406 (2007) / *cond-mat/0611542*.
[3] Y. Mnyukh, *J. Phys. Chem. Solids*, **24** (1963) 631.
[4] A.I. Kitaigorodskii, Y. Mnyukh, Y. Asadov, *Soviet Physics - Doclady* **8** (1963) 127.
[5] A.I. Kitaigorodskii, Y. Mnyukh, Y. Asadov, *J. Phys. Chem. Solids* **26** (1965) 463.
[6] Y. Mnyukh, N.N. Petropavlov, A.I. Kitaigorodskii, *Soviet Physics - Doclady* **11** (1966) 4.